\documentclass{ws-p9-75x6-50}

\usepackage{amsmath}
\usepackage{citesort}

\newcommand\newblock{}

\begin{document}

\title{New comparisons for local quantities \\ of the two-dimensional Hubbard
model}

\author{Adolfo Avella and Ferdinando Mancini}

\address{Dipartimento di Fisica ``E.R. Caianiello'' - Unit\`a INFM di Salerno\\
Universit\`a degli Studi di Salerno, I-84081 Baronissi
(SA),Italy\\ E-mail: avella@sa.infn.it}

\maketitle

\abstracts{We have compared the results of our approximation
scheme, the composite operator method, for the double occupancy
and the internal energy of the two-dimensional Hubbard model with
numerical data obtained by means of the Lanczos\cite{Becca:00} and
quantum Monte Carlo schemes\cite{White:89}. The agreement is very
good at both half-filling and away from it showing how reliable is
the approximation scheme.}

\section{Introduction}

The relevance of the Hubbard model\cite{Hubbard} as minimal model
for describing the high-T$_c$ cuprate
superconductors\cite{Anderson:87} and, generally, many strongly
correlated materials\cite{Fulde:95} is very well established. The
model is thought to have a very rich phenomenology: a variety of
spin and charge orderings\cite{Dagotto:94} (ferro- and antiferro-
magnetic included), Mott transitions (of the Slater, Heisenberg
and Hubbard types) driven by both filling and
coupling\cite{Gebhard:97,Avella:00b}, non-Fermi liquid dynamics,
superconductivity caused by: magnons, static and dynamic charge
ordering, proximity to quantum critical points of different
origin. In particular, strong antiferromagnetic correlations are
present at half-filling (and for low doping) and low
temperatures\cite{Dagotto:94}.

According to this, it is very relevant to have reliable solutions
of this model for the variety of boundary conditions under which
it can be studied. Actually, we know the exact solution only in
one dimension thanks to the Bethe Ansatz\cite{Lieb:68}. Some other
few exact results are known in quite special limits, but the more
relevant questions are still unsolved as the exact solution in two
dimensions (which seems to be the case for the majority of the
emergent highly interacting materials) is still missing. Owing to
this, a huge number of approximation schemes have been proposed
since the very beginning (Hubbard presented the model together
with an approximate solution known today as Hubbard I) and their
reliability is still under test as many physical interpretations
are based on the results obtained by them.

In the last decades we have seen the birth and development of many
numerical methods (exact diagonalization, Lanczos, quantum Monte
Carlo, ...) for studying finite clusters of bigger and bigger
size. The results of these numerical methods are extremely
important for the development of analytical schemes as they
provide the possibility to execute unbiased tests. The numerical
data can be interpreted as the experimental results relative to a
specific model and any comparison with an analytical approximation
scheme can be directly contrasted with its reliability. On the
contrary, comparisons with real experimental data suffer of an
high degree of uncertainty regarding the capability of the chosen
model to capture the essential physical of the material under
analysis.

In this last decade, we have been developing an approximation
scheme, namely the composite operator
method\cite{Mancini95,Avella:98,Mancini:98c,Avella:00b,Fiorentino:00a},
which proved to be quite powerful to describe the physics of many
correlated models and the properties of some materials. The main
peculiarities of this method are: the use of symmetry constraints
(Pauli principle and Ward-Takahashi identities) to fix the
representation where the Green's functions are realized, the
freedom in choosing the composite fields for the operatorial basis
around which we wish to construct perturbative solutions. The
algebra imposes many relations between operators which have to be
fulfilled also at the level of averages. According to this, it is
possible to individuate a set of self-consistent equations which
allows to fix the unknown parameters coming in the calculations
and, consequently, to instruct the solution regarding the Hilbert
space in which the composite operators live (i.e., the
representation). In particular, we see the unknown parameters as a
necessity and not as an accident as many other approximation
schemes do. In the last years, we have acquired much experience in
choosing the composite operators more effective in describing the
physical property of a system and we have collected a certain
number of recipes which can be used for specific class of models.

In this manuscript, we wish to present some new comparisons
between numerical results and composite operator method ones for
two local quantities: the double occupancy and the internal
energy.

\section{Results}

The Hubbard model is described by the following Hamiltonian
\begin{equation}
H=\sum_{\mathbf{ij}}\left( t_{\mathbf{ij}}-\mu \,\delta
_{\mathbf{ij}}\right) c^{\dagger }\left( i\right) \,c\left(
j\right) +U\sum_{\mathbf{i}}n_{\uparrow }\left( i\right)
\,n_{\downarrow }\left( i\right)
\end{equation}
where $c^{\dagger }\left( i\right) =\left( c_{\uparrow }^{\dagger
}\left( i\right) ,c_{\downarrow }^{\dagger }\left( i\right)
\right) $ is the electronic creation operator in spinorial
notation at the site $\mathbf{i}$ [$i=\left( \mathbf{i},t\right)
$] and $n_{\sigma }\left( i\right) =c_{\sigma }^{\dagger }\left(
i\right) \,c_{\sigma }\left( i\right) $ is the number operator for
spin $\sigma $ at the site $\mathbf{i}$; $\mu $ is the chemical
potential and $U$ is the on-site Coulomb repulsion.

The matrix $t_{\mathbf{ij}}$ describes the nearest-neighbor
hopping; in the 2D case we have $t_{\mathbf{ij}}=-4t\,\alpha
_{\mathbf{ij}}$, where
\begin{equation}
\alpha_{\mathbf{ij}}=\frac{1}{N}\sum_{\mathbf{k}}e^{\mathrm{i}\,\mathbf{k}(\mathbf{i}
-\mathbf{j})}\alpha \left( \mathbf{k}\right)
\end{equation}
is the projector on the nearest-neighbor sites and $\alpha \left(
\mathbf{k} \right) =\frac{1}{2}\left[ \cos \left( k_{x}\,a\right)
+\cos \left( k_{y}\,a\right) \right] $ and $a$ is the lattice
parameter.

We choose the following fermionic basis
\begin{equation}
\Psi \left( i\right) =\left(
\begin{array}{c}
\xi \left( i\right) \\
\eta \left( i\right)
\end{array}
\right)
\end{equation}
where $\xi \left( i\right) =\left[ 1-n\left( i\right) \right]
c\left( i\right) $ and $\eta \left( i\right) =n\left( i\right)
\,c\left( i\right) $ are the Hubbard operators. $\Psi \left(
i\right) $ satisfies the following equation of motion
\begin{equation}
J\left( i\right) =\mathrm{i}\frac{\partial }{\partial t}\Psi
\left( i\right) =\left(
\begin{array}{c}
-\mu \,\xi \left( i\right) -4t\,c^{\alpha }\left( i\right)
-4t\,\pi \left(
i\right) \\
-(\mu -U)\eta \left( i\right) +4t\,\pi \left( i\right)
\end{array}
\right)
\end{equation}
where $c^{\alpha}\left( \mathbf{i},t\right)
=\sum_{\mathbf{j}}\alpha_{\mathbf{ij}}\,c\left(
\mathbf{j},t\right)$ and $\pi \left( i\right) =\frac{1}{2}\sigma
^{\mu }\,n_{\mu }\left( i\right) \,c^{\alpha }\left( i\right)
+c\left( i\right) \left[ c^{\dagger \alpha }\left( i\right)
\,c\left( i\right) \right] $. $n_{\mu }(i)=c^{\dagger }(i)\,\sigma
_{\mu }\,c(i)$ are the charge ($\mu =0$) and spin ($\mu =1,2,3$ )
density operators, with $\sigma _{\mu }=\left(
1,\vec{\sigma}\right) $, $ \sigma ^{\mu }=\left(
-1,\vec{\sigma}\right) $ and $\vec{\sigma}$ are the Pauli
matrices.

Let us project the source $J\left( i\right)$ on the chosen basis
\begin{equation}
J\left( \mathbf{i},t\right) \cong \sum_{\mathbf{j}}\varepsilon
\left( \mathbf{i},\mathbf{j}\right) \,\Psi \left(
\mathbf{j},t\right) \label{psipem}
\end{equation}
Accordingly, the energy matrix $\varepsilon \left(
\mathbf{i},\mathbf{j}\right) $ is defined through the equation
\begin{equation}
m\left( \mathbf{i},\mathbf{j}\right) =\sum_{\mathbf{l}}\varepsilon
\left( \mathbf{i},\mathbf{l}\right) \,I\left(
\mathbf{l},\mathbf{j}\right)
\end{equation}
where the $m$-matrix and the normalization matrix $I$ have the
following definitions
\begin{align}
&m\left( \mathbf{i},\mathbf{j}\right) =\left\langle \left\{
J\left( \mathbf{i},t\right) ,\Psi
^{\dagger }\left( \mathbf{j},t\right) \right\} \right\rangle \\
&I\left( \mathbf{i},\mathbf{j}\right) =\left\langle \left\{ \Psi
\left( \mathbf{i},t\right) ,\Psi ^{\dagger }\left(
\mathbf{j},t\right) \right\} \right\rangle
\end{align}
After Eq.~(\ref{psipem}), the Fourier transform of the thermal
single-particle retarded Green's function $G\left( i,j\right)
=\langle R\left[ \Psi \left( i\right) \,\Psi ^{\dagger }\left(
j\right) \right] \rangle $ satisfies the following equation
\begin{equation}
\left[ \omega -\varepsilon \left( \mathbf{k}\right) \right]
G\left( \mathbf{k },\omega\right) =I\left( \mathbf{k}\right)
\end{equation}

The straightforward application of this
scheme\cite{Mancini95,Avella:98,Mancini:98c,Avella:00b,Fiorentino:00a}
gives that, in the paramagnetic phase, $I\left( \mathbf{k}\right)
$ has diagonal form with $I_{11}=1-n/2$ and $I_{22}=n/2$ ($
\langle n_{\sigma }\left( i\right) \rangle =\frac{n}{2}$) and that
the $m$-matrix depends on three parameters: the chemical potential
$\mu $ and two correlators
\begin{align}
&\Delta = \langle \xi ^{\alpha }\left( i\right) \,\xi ^{\dagger
}\left( i\right) \rangle -\langle \eta ^{\alpha }\left( i\right)
\,\eta ^{\dagger
}\left( i\right) \rangle \\
&p = \frac14\langle n_{\mu }^{\alpha }\left( i\right) \,n_{\mu
}\left( i\right) \rangle -\langle \lbrack c_{\uparrow }\left(
i\right) \,c_{\downarrow }\left( i\right) ]^{\alpha }c_{\downarrow
}^{\dagger }\left( i\right) \,c_{\uparrow }^{\dagger }\left(
i\right) \rangle
\end{align}
The three parameters $\mu $, $\Delta $ and $p$ are functions of
the external parameters $n$, $T$ and $U$ and can be determined
self-consistently through a set of three coupled equations
\begin{equation}
\left\{
\begin{array}{l}
n=2\left[ 1-\left\langle c\left( i\right) \,c^{\dagger }\left(
i\right)
\right\rangle \right] \\
\Delta =\langle \xi ^{\alpha }\left( i\right) \,\xi ^{\dagger
}\left( i\right) \rangle -\langle \eta ^{\alpha }\left( i\right)
\,\eta ^{\dagger
}\left( i\right) \rangle \\
\langle \xi \left( i\right) \,\eta ^{\dagger }\left( i\right)
\rangle =0
\end{array}
\right.
\end{equation}
The first equation fixes the particle number, the second one comes
from the definition of $\Delta $ and the third one assures that
the solution respects the Pauli principle (i.e., the local
algebra)\cite{Mancini:00}.

\begin{figure}[t]
\center{\epsfxsize=8cm \epsfbox{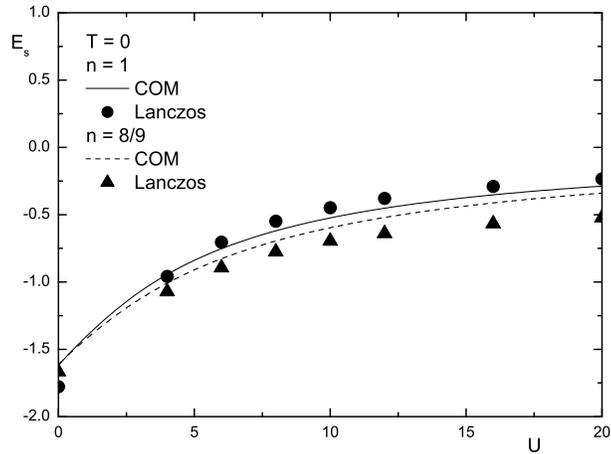} \caption{Internal energy
per site $E_s$ as a function of the Coulomb interaction $U$ for
$T=0$ and $n=1$ and $8/9$. The Lanczos data are taken from
Ref.~\protect\cite{Becca:00}. \label{Fig1}}}
\end{figure}

\begin{figure}[t]
\center{\epsfxsize=8cm \epsfbox{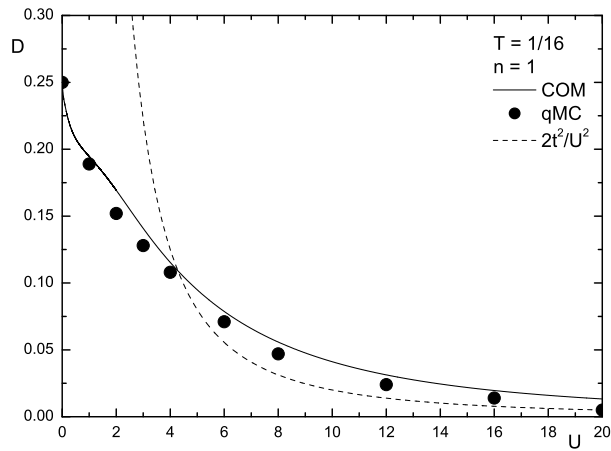} \caption{Double
occupancy $D$ as a function of the Coulomb interaction $U$ for
$T=1/16$ and $n=1$. The quantum Monte Carlo data are taken from
Ref.~\protect\cite{White:89}. \label{Fig2}}}
\end{figure}

\begin{figure}[t]
\center{\epsfxsize=8cm \epsfbox{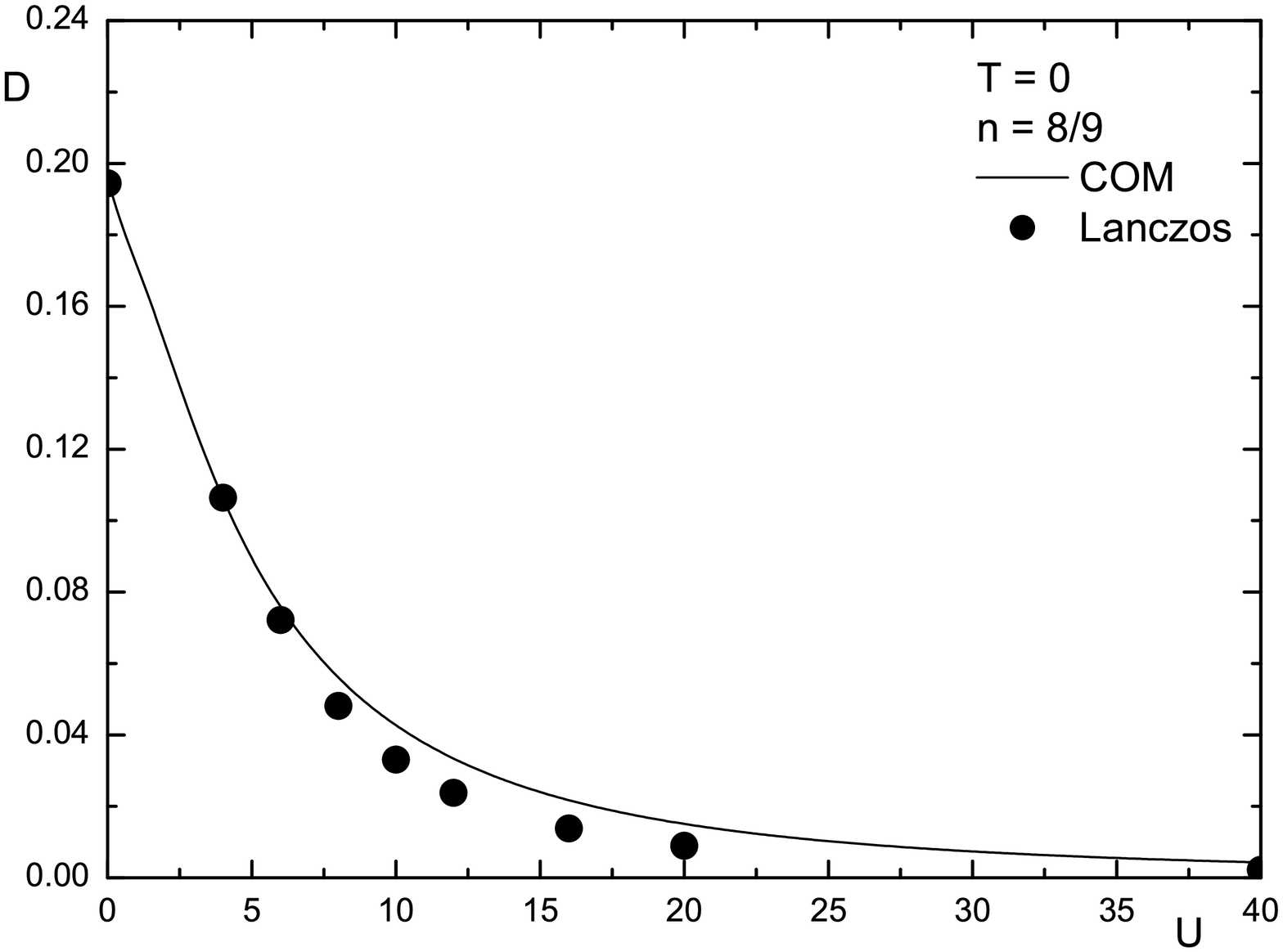} \caption{Double
occupancy $D$ as a function of the Coulomb interaction $U$ for
$T=0$ and $n=8/9$. The Lanczos data are taken from
Ref.~\protect\cite{Becca:00}. \label{Fig3}}}
\end{figure}

In Fig.~\ref{Fig1} we report the behavior of the internal energy
per site $E_s=\langle H \rangle/N$ as a function of the Coulomb
interaction $U$ at half-filling and $n=8/9$. The Lanczos numerical
data have been taken from Ref.~\protect\cite{Becca:00}. The
agreement at half-filling is very good. We have got an
exceptionally good agreement at half-filling for the
one-dimensional Hubbard model too\cite{Avella:98e}. In that case,
we have compared our results with the exact Bethe ansatz solution.
Away from half-filling, the agreement is less satisfactory, but,
as we will see from the double occupancy comparison, the
difference is quite constant as function of $U$ and can be due to
finite size effects, as surely is the difference at $U=0$.

The double occupancy $D=\langle n_\uparrow n_\downarrow \rangle$,
which can be computed by means of thermodynamics through the
following formula $D=\frac{\partial F}{\partial U}$ ($F$ is the
free energy; for $T=0$ we have $D=\frac{\partial E_s}{\partial
U}$), is reported as a function of the Coulomb interaction $U$ in
Fig.~\ref{Fig2} for $T=1/16$ and $n=1$ and in Fig.~\ref{Fig3} for
$T=0$ and $n=8/9$. The Lanczos numerical data have again been
taken from Ref.~\protect\cite{Becca:00}; the quantum Monte Carlo
ones from Ref.~\protect\cite{White:89}. The agreement is very good
for both values of filling. For half-filling, we have also
reported the limiting behavior of our solution: $D\stackrel{U
\rightarrow \infty}{\longrightarrow} d \frac{t^2}{U^2}$ where $d$
is the spatial dimension of the system\cite{Mancini:00a}. The
numerical data seem to follow this behavior also for not so large
value of the Coulomb repulsion. This is probably due to finite
size effects. For $n=8/9$, the very good agreement supports our
interpretation of the comparison for the corresponding energy.

\section{Conclusions}

The positive comparisons with the Bethe ansatz exact solution for
the one-dimensional case and with the numerical results for the
two-dimensional one show how reliable is our approximation scheme
to study the Hubbard model. In particular, the agreement reported
in this manuscript with the Lanczos and the quantum Monte Carlo
data at both half-filling and away from it is very good and shows
our capability to properly describe both the charge and the spin
fluctuation scales of energy.

\end{document}